\documentclass[linenumbers,twocolumn]{aastex631}
\usepackage{listings}
\usepackage{xcolor}
\usepackage[utf8]{inputenc}  
\usepackage{csvsimple}
\usepackage{booktabs}
\usepackage{pgfplotstable}
\usepackage{datatool}
\usepackage{hyperref}

\definecolor{Dkgreen}{rgb}{0,0.6,0}
\definecolor{Gray}{rgb}{0.5,0.5,0.5}
\definecolor{Mauve}{rgb}{0.58,0,0.82}
\definecolor{Red}{rgb}{1,0,0}
\definecolor{Violet}{rgb}{0.93,0.51,0.93}
\definecolor{Blue}{rgb}{0,0,1}

\pgfplotsset{compat=1.18}
\begin{document}
\nolinenumbers
\title{Can we constrain warm dark matter masses with individual galaxies?}

\author[0009-0000-5381-7039]{Shurui Lin}
\affiliation{Department of Astronomy, School of Physical Sciences, University of Science and Technology of China, Hefei, Anhui 230026, China}
\affiliation{CAS Key Laboratory for Researches in Galaxies and Cosmology, School of Astronomy and Space Science, University of Science and Technology of China, Hefei, Anhui 230026, China}

\author[0000-0002-4816-0455]{Francisco Villaescusa-Navarro}
\affiliation{Center for Computational Astrophysics, 162 5th Avenue, New York, NY, 10010, USA}
\affiliation{Department of Astrophysical Sciences, Princeton University, 4 Ivy Lane, Princeton, 
NJ 08544 USA}

\author[0000-0002-2628-0237]{Jonah Rose}
\affiliation{Department of Astronomy, University of Florida, Gainesville, FL 32611, USA}
\affiliation{Center for Computational Astrophysics, 162 5th Avenue, New York, NY, 10010, USA}

\author[0000-0002-5653-0786]{Paul Torrey}
\affiliation{Department of Astronomy, University of Virginia, 530 McCormick Road, Charlottesville, VA 22903, USA}

\author[0000-0003-0777-4618]{Arya~Farahi}
\affiliation{Departments of Statistics and Data Science, University of Texas at Austin, Austin, TX 78757, USA}

\author[0000-0003-4004-2451]{Kassidy E. Kollmann}
\affiliation{Department of Physics, Princeton University, Princeton, NJ 08544, USA}

\author[0000-0002-8111-9884]{Alex M. Garcia}
\affiliation{Department of Astronomy, University of Virginia, 530 McCormick Road, Charlottesville, VA 22903, USA}

\author[0000-0002-7638-7454]{Sandip Roy}
\affiliation{Department of Physics, Princeton University, Princeton, NJ 08544, USA}

\author[0000-0002-3204-1742]{Nitya Kallivayalil}
\affiliation{Department of Astronomy, University of Virginia, 530 McCormick Road, Charlottesville, VA 22904 USA}

\author[0000-0001-8593-7692]{Mark Vogelsberger}
\affiliation{Department of Physics and Kavli Institute for Astrophysics and Space Research, Massachusetts Institute of Technology, Cambridge, MA 02139, USA}
\affiliation{The NSF AI Institute for Artificial Intelligence and Fundamental Interactions, Massachusetts Institute of Technology, Cambridge, MA 02139, USA}

\author[0000-0003-0706-8465]{Yi-Fu Cai}
\affiliation{Department of Astronomy, School of Physical Sciences, University of Science and Technology of China, Hefei, Anhui 230026, China}
\affiliation{CAS Key Laboratory for Researches in Galaxies and Cosmology, School of Astronomy and Space Science, University of Science and Technology of China, Hefei, Anhui 230026, China}

\author[0000-0003-1297-6142]{Wentao Luo}
\affiliation{Department of Astronomy, School of Physical Sciences, University of Science and Technology of China, Hefei, Anhui 230026, China}

\author{et al.}

\begin{abstract}
\nolinenumbers
We study the impact of warm dark matter mass on the internal properties of individual galaxies using a large suite of 1,024 state-of-the-art cosmological hydrodynamic simulations from the DREAMS project. We take individual galaxies' properties from the simulations, which have different cosmologies, astrophysics, and warm dark matter masses, and train normalizing flows to learn the posterior of the parameters. We find that our models cannot infer the value of the warm dark matter mass, even when the values of the cosmological and astrophysical parameters are given explicitly. This result holds for galaxies with stellar mass larger than $2\times10^8 M_\odot/h$ at both low and high redshifts. We calculate the mutual information and find no significant dependence between the WDM mass and galaxy properties. On the other hand, our models can infer the value of $\Omega_{\rm m}$ with a $\sim10\%$ accuracy from the properties of individual galaxies while marginalizing astrophysics and warm dark matter masses.
\end{abstract}

\keywords{}

\section{Introduction} \label{sec:intro}

Recent works have identified the existence of a tight correlation between the fraction of matter in the total matter density of our universe $\Omega_{\rm m}$ and the properties of individual galaxies in state-of-the-art hydrodynamic simulations \citep{Paco2022APJ, Echeverri, Chaitanya}.

By training neural networks on simulations that implement various cosmologies and astrophysical models, the authors found they could infer the value of $\Omega_{\rm m}$ with a $\sim10\%$ precision using the properties of individual galaxies while marginalizing over astrophysics.
They also concluded that stellar mass, stellar metallicity, and maximum circular velocity are the most important properties for the inference.
This idea has been used with observational data to derive the first constraints on cosmological parameters using just the photometry of galaxies by \cite{Chang_2023}.

\cite{Paco2022APJ} indicated that the existence of the above correlation may be due to the fact that changing the value of $\Omega_{\rm m}$ affects the overall dark matter content of galaxies and their assembly history, changing internal galaxy properties in a way that is not degenerate with feedback. In other words, injecting dark matter into galaxies (or changing their assembly history) has an impact on galaxy properties that is different (when looking at the data in high dimensions) from the one induced by feedback \citep{Echeverri,Chaitanya}. 

In this work, we ask ourselves: can we use the properties of individual galaxies to constrain the nature of dark matter? To test the ability of single galaxy properties to infer dark matter properties, we use the simplest possible alternative dark matter model: Warm Dark Matter (WDM). In contrast to Cold Dark Matter (CDM), our WDM model is characterized by a single parameter that describes the mass of the DM particle (or, alternatively, the cutoff scale in the initial matter power spectrum).  Using a similar simulation suite to what we explore here (containing N-body simulations only), \cite{Jonah2023MNRAS} previously showed that convolutional neural networks — trained on the matter density field — could efficiently infer WDM properties with higher accuracy than employing standard summary statistics such as the matter power spectrum. In this work, our study contains two key changes from that of \cite{Jonah2023MNRAS}.  First, in place of the dark matter only simulations used in \cite{Jonah2023MNRAS}, we employ a suite of simulations that includes a full treatment of baryons based on the IllustrisTNG model. This means that, in addition to dark matter properties, we also have important information about the baryon content of galaxies. Second, instead of using field-level inference based on the full matter density field, we focus on individual galaxy properties this time.

To quantify the dependence of galaxy properties on WDM mass, we focus our attention on the posterior distribution $p(\vec{\theta}|\mathbf{X})$, where $\mathbf{X}$ are the properties of individual galaxies and $\vec{\theta}$ are the cosmological, astrophysical, and WDM mass parameters. What we require is a method to model the multi-dimensional joint posterior of $\vec{\theta}$ parameters. Normalizing flows meet our needs perfectly as they can offer exact posterior modeling and efficient sampling \citep{NF2016variational,NF2021JMLR}. We train our model on a dataset that consists of hundreds of thousands of galaxies from 1,024 state-of-the-art cosmological hydrodynamic simulations of the DaRk mattEr with AI and siMulationS (DREAMS) project. Each simulation implements a different cosmology, astrophysics, and warm dark matter mass. We train our models using both galaxies at high-redshift ($z=5$), where the effect of WDM is expected to be higher, and low-redshift ($z=0$). Overall, we find negligible dependence of the properties of individual galaxies on WDM masses.

This paper is organized as follows. In Section \ref{sec:methods} we present the hydrodynamic simulations used in this work, the properties of the galaxies employed, and machine learning methods used to train and test the models. We then show the main results of this work in Section \ref{sec:results}. Finally, we discuss our findings and draw conclusions in Section \ref{sec:conclusions}.

%%%%%%%%%%%%%%%%%%%%%%%%%%%%%%%%%%%%%%%%%%%%%%%%%%%%%%
%%%%%%%%%%%%%%%%%%%%%%%%%%%%%%%%%%%%%%%%%%%%%%%%%%%%%%
\section{Methods} 
\label{sec:methods}

In this section, we describe the data we use and the machine learning models we utilize to learn the posterior distribution.

\subsection{Simulations} 
\label{subsec: simulations}

We made use of a suite of 1,024 magneto-hydrodynamic cosmological simulations from the DaRk mattEr with AI and siMulationS (DREAMS) project \citep{darkCAMELS}. Each simulation follows the evolution of $256^3$ dark matter particles plus $256^3$ initial fluid elements from the initial conditions, generated at $z=127$ using second-order Lagrangian perturbation theory (2LPT), down to $z=0$ in a comoving volume of $(25~h^{-1}{\rm Mpc})^3$. The dark matter and gas\footnote{By gas mass resolution, we mean the masses of the gas particles in the initial conditions. We note that the masses of the gas and star particles can vary during the time evolution of the simulation.} mass resolution is $6.44\times10^7 \times ((\Omega_\mathrm{m}$-0.049) / 0.251) $h^{-1}M_\odot$ and 1.3 $\times$ 10$^7$ $h^{-1}M_\odot$, respectively, while the spatial resolution is 1.0 kpc at $z=0$. 

The simulations are run with the moving mesh code \textsc{arepo} \citep{Springel2010, Springel2019, Weinberger2020} using the IllustrisTNG galaxy formation model \citep{2018Pillepich, 2018Weinberger}.
The IllustrisTNG models have been well-tested across a range of resolutions and environments and have been shown to match many galaxy scaling relations and observables \citep{2018Pillepichb,2018Springel}.

All simulations use the same value for the following cosmological parameters $\Omega_{\rm b}=0.049$, $h=0.6711$, $n_s=0.9624$, $M_\nu=0$ eV, $\Omega_{\rm K}=0$ while each simulation implements a different value for:

\begin{itemize}
\item the initial random seed
\item $\Omega_{\rm m}$ and $\sigma_8$
\item three astrophysical parameters that control the efficiency of feedback from supernovae and AGN: $A_{\rm SN1}$, $A_{\rm SN2}$, $A_{\rm AGN1}$
\item the warm dark matter mass
\end{itemize}

We now briefly describe the meaning of the astrophysical parameters. Galactic winds induced by supernova feedback in the IllustrisTNG model are modeled by the \cite{2003Springel} model. As implemented in \textsc{arepo}, this model includes two normalization factors that control the energy injection rate and the wind speed. Those normalization parameters represent our $A_{\rm SN1}$ and $A_{\rm SN2}$ free parameters\footnote{We note that these two parameters are also varied in the CAMELS IllustrisTNG simulations. See \cite{CAMELS} for more details.}.
The AGN parameter, $A_{\rm AGN1}$, controls the normalization factor of AGN feedback in the high-accretion state.
We refer the reader to \cite{darkCAMELS} for more information and a discussion on why this parameter was changed from those in the CAMELS simulations.

The value of the cosmological, astrophysical, and warm dark matter mass parameters are arranged in a Sobol sequence (a special kind of low-discrepancy sequence for random sampling) \citep{sobol1967} with very broad boundaries defined by:
\begin{equation}
0.1\leq\Omega_{\rm m}\leq0.5
\end{equation}
\begin{equation}
0.6\leq \sigma_8\leq1.0
\end{equation}
\begin{equation}
0.25\leq A_{\rm SN1}\leq4
\end{equation}
\begin{equation}
0.5\leq A_{\rm SN2} \leq 2
\end{equation}
\begin{equation}
0.25\leq A_{\rm AGN1} \leq 4
\end{equation}
\begin{equation}
    1.8~{\rm keV} \leq M_{\rm wdm} \leq 16~ {\rm keV}
\end{equation}
We adopt broad sampling to avoid the results being dominated by tight priors that can lead us to the wrong conclusions. This approach will ensure a conservative marginalization over baryonic effects.
We note that the values of $\Omega_{\rm m}$ and $\sigma_8$ are sampled linearly, while the value of the astrophysical parameters are sampled logarithmically. The warm dark matter masses are sampled uniformly in ${\rm keV}/M_{\rm wdm}$. We note that the center of the Sobol sequence for the astrophysical parameter corresponds to the fiducial IllustrisTNG model.

In this work, we assume that warm dark matter is produced thermally. Thus, its velocity distribution is given by the Fermi-Dirac distribution and can be fully characterized by its width, which in turn, depends on its mass. This effect induces a change in the initial matter power spectrum on small scales, which is characterized by the following transfer function $\beta(k)$, P$_\mathrm{wdm}(k)$ = $\beta(k)^2$P$_\mathrm{cdm}(k)$, where
\begin{equation}
    \beta(k) = \left( 1 + (k \alpha)^{2.4} \right)^{-5.0/1.2}
\end{equation}
\begin{equation}
        \alpha = 0.049 \left( \frac{M_{\mathrm{wdm}}}{1~\mathrm{keV}} \right)^{-1.11} \left( \frac{\Omega_m - \Omega_b}{0.25} \right)^{0.11} \left( \frac{h}{0.7} \right)^{1.22}
\end{equation}
with $k$ being the wavenumber, $\Omega_m$ ($\Omega_b$) is the total matter (baryonic) density of the universe, and $h$ is the reduced Hubble constant. We refer the reader to \cite{Bode2001APJ, Jonah2023MNRAS} for further details on this. The WDM mass boundaries (1.8 keV and 16 keV) correspond to a 99.4\% and 3.0\% reduction, respectively, in the initial power spectrum at our Nyquist frequency (32 $h$ Mpc$^{-1}$).

\subsection{Galaxy properties}
\label{subsuc: gal_prop}

We identify halos and subhalos in the simulations using SUBFIND \citep{Springel2001MNRAS}. In this work, we define galaxies as subhalos with a stellar mass larger than $2\times10^8~h^{-1}M_\odot$, regardless of whether the galaxy is a central or a satellite. Each simulated galaxy is characterized by many different properties, but in this work, we focus our attention on the following 14:

\begin{enumerate}
    \boldmath
    \item $M_\mathrm{g}$. 
    The total mass of the gas content in the galaxy.
    \item $M_\mathrm{BH}$. 
    The total mass of black holes in the galaxy.
    \item $M_\mathrm{*}$. 
    The total mass of stellar components in the galaxy.
    \item $M_\mathrm{t}$.
    The total mass of the halo hosting the galaxy, 
    including the mass of dark matter, gas, stars, and black holes in the all subhalo.
    \item $V_\mathrm{max}$.
    \unboldmath
    The highest circular velocity within the subhalo hosting the galaxy: $V_\mathrm{max} = \mathrm{max}(\sqrt{GM(<R)}/R)$
    \boldmath
    \item $\sigma_v$.
    The norm of velocity dispersion of all member particles of the subhalo.
    \item $Z_\mathrm{g}$.
    The mass-weighted metallicity of the gas within the galaxy.
    \item $Z_\mathrm{*}$.
    The mass-weighted metallicity of the stellar components of the galaxy.
    \item \textbf{SFR}.
    Star-formation rate of the galaxy.
    \item $J$.
    The modulus of the spin vector for the galaxy's subhalo.
    \item $V$.
    The modulus of peculiar velocity for the galaxy's subhalo.
    \item $R_\mathrm{*}$ .
    The radius encompassing 50\% of the galaxy's entire stellar mass.
    \item $R_\mathrm{t}$.
    The radius encompassing 50\% of the galaxy's total mass.
    \item $R_\mathrm{max}$.
    \unboldmath
    The radius where $GM(< R_\mathrm{max})/R_\mathrm{max} = V_\mathrm{max}$.

\end{enumerate}

We work with two different galaxy populations:
\begin{itemize}
\item All galaxies: $M_*>2\times10^8~h^{-1}M_\odot$
\item Small galaxies: $2 \times 10^8~ h^{-1}M_\odot \leq M_* \leq 10^9~h^{-1}M_\odot$
\end{itemize}

\subsection{Data preprocessing}
\label{subsec:preprocessing}

Our model will be built by passing to it pairs $\{\mathbf{X},\vec{\theta}\}$, where $\mathbf{X}$ are some input features (the 14 properties of the individual galaxies) and $\vec{\theta}$ is the galaxy parameters (e.g. the value of $\Omega_{\rm m}$ and the WDM mass). 

Having galaxies from 1,024 simulations, we first split the simulations into training (80\%), validation (10\%), and testing (10\%).
We split the simulations (not the individual galaxies) to group all galaxies from a given simulation into one dataset. Splitting in this way avoids any potential leakage of information.

This is because galaxies in the same simulation may share some properties given to internal correlations. 

As the number of galaxies in each simulation is correlated to its value of $\Omega_\mathrm{m}$ and $\sigma_8$, we randomly select 500 galaxies for every simulation to avoid bias.\footnote{We do this selection to avoid bias. Another choice of number (like 200 galaxies per simulation) does not affect our inference significantly.} In the cases where the simulation does not contain that many galaxies, we randomly repeat some until meeting that threshold.

We then normalize every dimension for each galaxy by substracting the mean and dividing by the standard deviation. We first take the log before the normalization for all 14 features listed above. We follow a similar procedure for the cosmological and astrophysical parameters, without taking the log. For the WDM mass parameter, we use its inverse, which is linearly sampled.

\subsection{Machine Learning algorithms}
\label{subsec:ML_alg}

Our goal in this paper is to learn the posterior distribution $p(\vec{\theta}|\mathbf{X})$, where $\mathbf{X}$ represent properties of individual galaxies and $\vec{\theta}$ represent the cosmological, astrophysical, and/or WDM mass parameters of the corresponding galaxy using normalizing flows. 

Normalizing flows are models that map data from a base distribution (typically a multivariant Gaussian) to another (typically more complex) distribution \citep{NF2021JMLR}. Let's consider a random vector drawn from the base distribution $\mathbf{u}\sim \pi(\mathbf{u})$. If we have an invertible and differentiable function $f_\phi$, where $\phi$ are some parameters, then the variable $\mathbf{x}=f_\phi(\mathbf{u})$ will have a distribution $\mathbf{x}\sim p(\mathbf{x})$ given by 
\begin{equation}
    p(\mathbf{x}) = \pi\left(f^{-1}_\phi(\mathbf{x})\right)
    \left|
    \det\left(\frac{\partial f^{-1}_\phi}{\partial \mathbf{x}}\right)
    \right|
\end{equation}

It can be shown with enough flexibility in the function $f_\phi$, the above method can be used to learn any generic distribution. In our particular case, we made use of a spline autoregressive flow. In these models, the target distribution $p(\mathbf{x})$ is decomposed as the products of 1D pdfs
\begin{equation}
p(\mathbf{x})=\prod_{i=1}^D p(x_i|x_{1;i-1})~,
\end{equation}
where $x_{1;i-1}=[x_1, x_2, ..., x_{i-1}]$. Each individual conditional pdf is then modeled as a parametric density that depends on some state $h_i$, which depends on $h_{i-1}$ and $x_i$. In the case of the neuro spline, $p(x_i|x_{1;i-1})$ is modeled by splines whose parameters are predicted by neural networks. We refer the reader to \cite{Durkan2019arXiv,Dolatabadi2020arXiv} for further details on these models.

The architecture of our model consists of several layers, within which there is a linear rational spline bijection coupled to an autoregressive neural network layer consisting of two hidden layers with Rectified Linear Unit (ReLU) as an activation function. We use the Pyro\footnote{\url{https://docs.pyro.ai/en/stable/index.html}} package to train, validate, and test our model. We train the model by passing pairs $\{\mathbf{X},\vec{\theta}\}$ and maximizing the log-likelihood of the model shown in  Eq. \ref{eq:loss} by tuning its free parameters with gradient descent using the Adam optimizer.
\begin{equation}
    \mathcal{L} = -\mathrm{log}[p(\vec{\theta}|\mathbf{X})]
\label{eq:loss}
\end{equation}

The hyperparameters of the model are: 1) the number of layers, 2) the number of spline segments, 3) the number of hidden dimensions for the autoregressive neural networks, 4) the bounding box of the spline\footnote{Note that this parameter does not need to be optimized if the data is normalized in a particular way. In our case, for simplicity, we decided to optimize it.}, 5) the learning rate, and 6) the weight decay. We use the Optuna\footnote{\href{https://optuna.org}{https://optuna.org}} package \citep{Akiba2019DBLP} to perform the hyperparameter optimization, by using 200 trials. During hyperparameter tuning, we search for the value of the hyperparameters that minimize the validation loss. In Table \ref{table:hpt}, we show the range we used for the different parameters for Optuna.

\begin{table}
  \centering
  \begin{tabular}{|c|c|c|}
    \hline
    Hyperparameters & Min & Max \\
    \hline
    Number of layers & 1 & 10 \\
    Spline segments & $2^1$ & $2^6$ \\
    NN hidden dimensions & $2^4$ & $2^8$ \\
    Bound & 1 & 10 \\
    Learning rate & $10^{-5}$ & $10^{-2}$\\
    Weight decay & $10^{-5}$ & $10^{-1}$\\
    \hline
  \end{tabular}
    \caption{This table shows the hyperparameters considered and their range. Hyperparameters are tuned by running Optuna for 200 trials.}
    \label{table:hpt}
\end{table}

\begin{figure*}
\includegraphics[width=2\columnwidth]{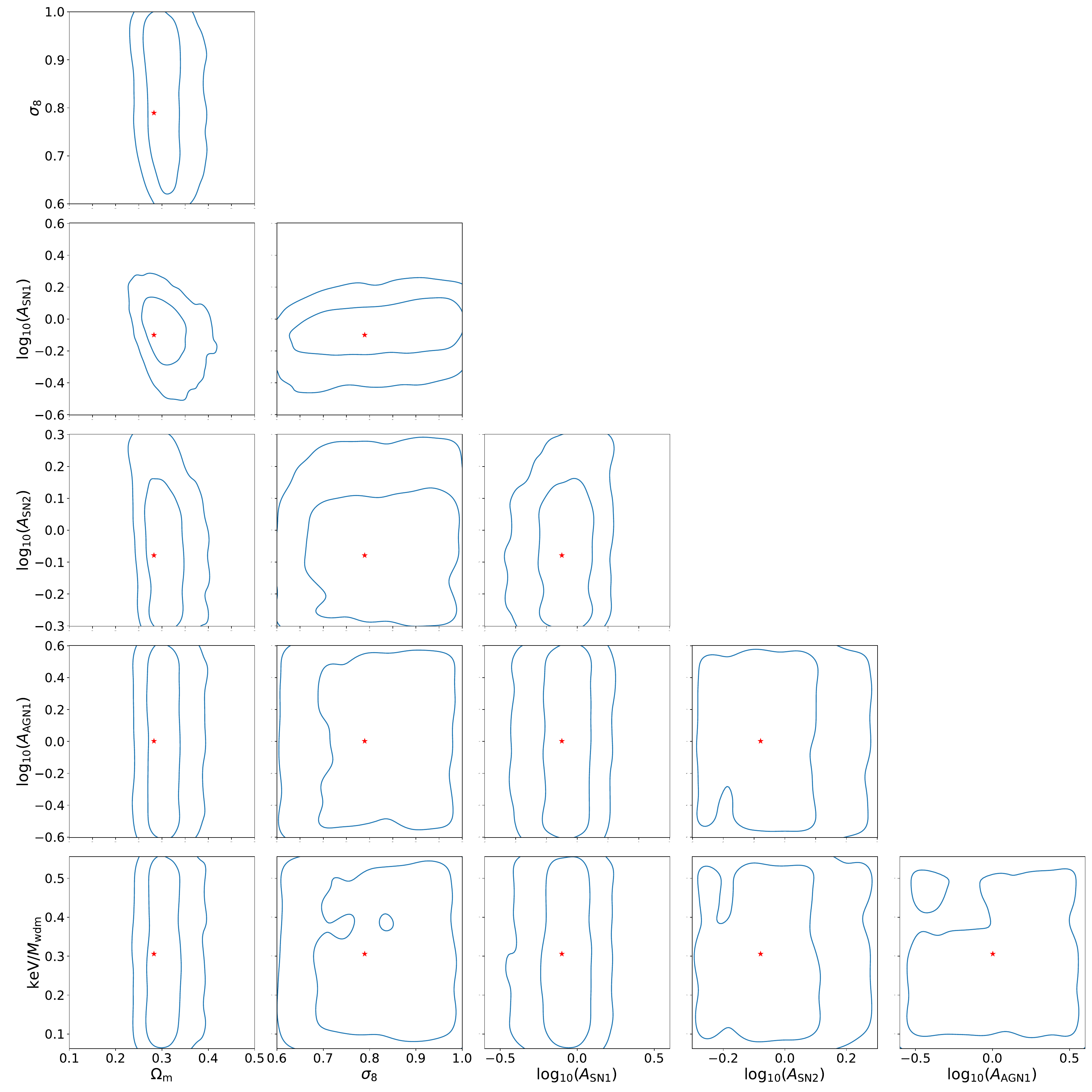}
\caption{With our normalizing flow trained using galaxy properties at $z=5$, we take a random galaxy of the test set and generate 10000 samples of the posterior given its properties. The red star indicates the true value of the parameters, while the blue contours indicate the 68\% and 95\% confidence intervals. As we can see, our model gives quite good predictions on $\Omega_\mathrm{m}$, while having poor performance on the others.}
\label{fig:contour}
\end{figure*}

\begin{figure*}
\includegraphics[width=2\columnwidth]{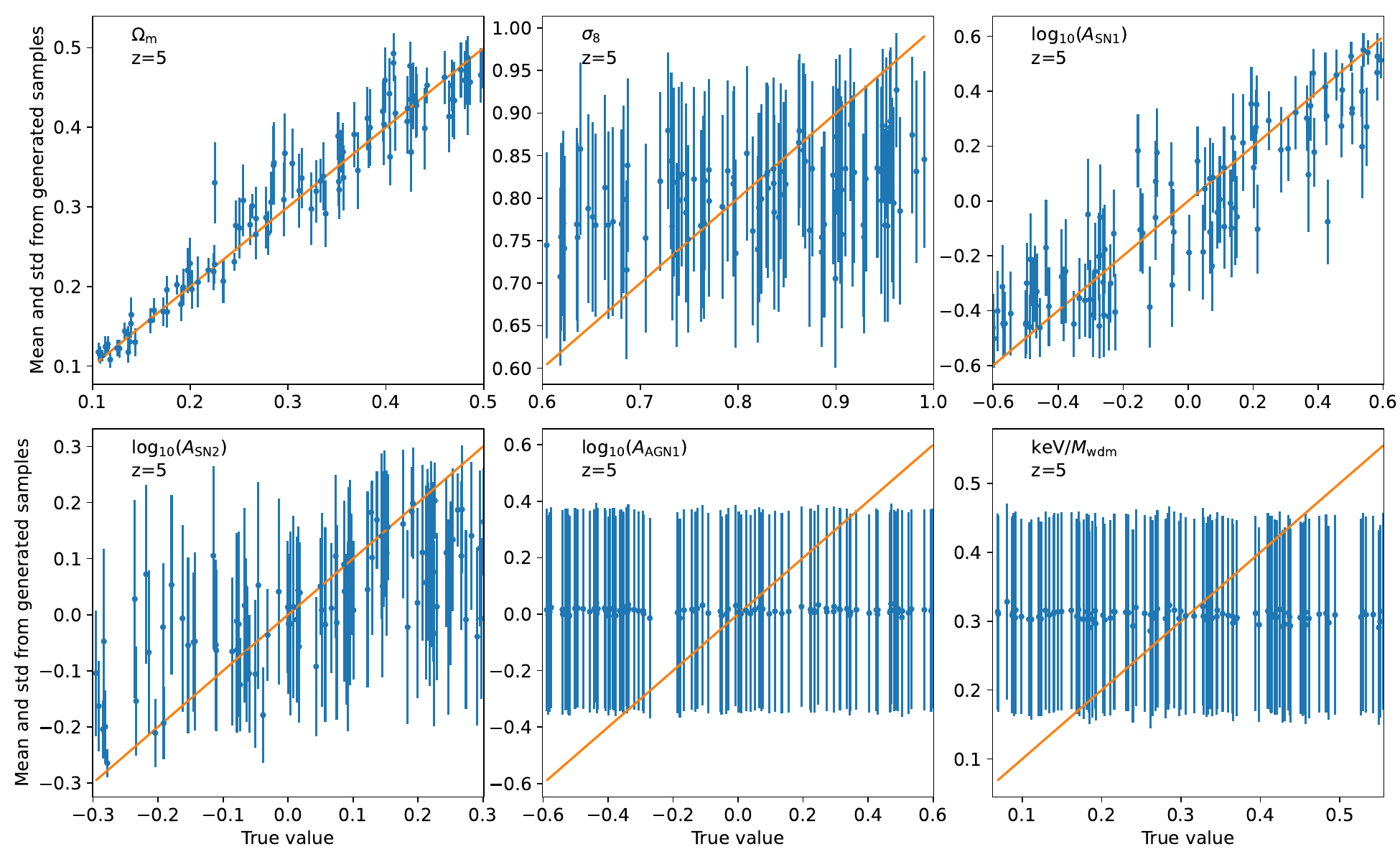}
\caption{We have trained a conditional normalizing flow to learn the posterior $p(\vec{\theta}|\mathbf{X})$, where $\mathbf{X}$ are 14 properties of single galaxies at $z=5$ and $\vec{\theta}$ are the values of the cosmological, astrophysical, and WDM mass parameters of the galaxy. Once the model is trained, we take a random set of galaxies from the test set and draw samples of the posterior from it. We then compute, for every galaxy, the posterior mean and the standard deviation from the samples and show the results in the different panels. For WDM mass, we have $\rm RMSE=0.1411$, $\epsilon=46.2\%$ and $R^2 = -0.0351$, which means our model is not able to infer the WDM mass of the galaxy. However, for $\Omega_{\rm m}$, our model can get the value acurrately with $\rm RMSE=0.0301$, $\epsilon=10.2\%$ and $R^2 = 0.9372$.}
\label{fig:6_properties}
\end{figure*}

\subsection{Validation metrics}
\label{subsec:st_var}

We quantify how well our model has learned the underlying distribution by performing coverage tests. Furthermore, we quantify the model's accuracy and precision using three simple metrics, listed below.
For the coverage test, we made use of the ``Tests
of Accuracy with Random Points'' (TARP) method described in \cite{lemos2023arXiv} using 2048 samples. 

Once the model is trained, it has approximated the posterior distribution $p(\vec{\theta}|\mathbf{X})$. Thus, given an input, one can generate samples from the distribution. We use these samples to:
\begin{itemize}
\item perform coverage tests for all parameters and each parameter individually
\item estimate simple accuracy and precision metrics for each parameter individually
\end{itemize}

 For this, we generate $10^6$ samples $\theta_i$ and estimate the first two moments of the distribution using:
\begin{equation}
    \mu_i = \frac{1}{N}\sum_i \theta_i
\end{equation}
\begin{equation}
    \sigma_i^2 = \frac{1}{N} \sum_i (x_i -\mu_i)^2
\end{equation}

From these quantities, we can calculate some validation metrics\footnote{We note that there are more powerful statistics tests than the ones presented here, but we employ them as a way to compare with previous works \citep{Paco2022APJ, Echeverri, Chaitanya} that used these metrics due to the different inference method.}:

\begin{enumerate}
\item \textbf{Root Mean Squared Error (RMSE)}
\begin{equation}
    \mathbf{ RMSE} = \sqrt{
    \langle(\theta_i - \mu_i)^2\rangle
    }
\end{equation}
Smaller RMSE implies a more accurate result.
\item \textbf{Mean Relative Error ($\epsilon$)}
\begin{equation}
    \epsilon = 
    \left\langle\frac{\sigma_i}{\mu_i}\right\rangle
\end{equation}
A smaller value means a more precise result.
\item \textbf{Coefficient of Determination ($R^2$)}
\begin{equation}
    R^2 = 1- 
    \left\langle
    \frac{(\theta_i - \mu_i)^2}
    {(\theta_i - \Bar{\theta}_i)^2}
    \right\rangle
\end{equation}
The smaller the difference between $R^2$ and 1, the more accurate the model is.
\end{enumerate}

%%%%%%%%%%%%%%%%%%%%%%%%%%%%%%%%%%%%%%%%%%%%%
%%%%%%%%%%%%%%%%%%%%%%%%%%%%%%%%%%%%%%%%%%%%%
\section{Results} 
\label{sec:results}

We now show the results we obtain after training our models.

\subsection{Parameter inference}
\label{subsec:all_gal}

We first train our models to learn $p(\vec{\theta}|\mathbf{X})$, where $\vec{\theta}=\{\Omega_{\rm m}, \sigma_8, A_{\rm SN1}, A_{\rm SN2}, A_{\rm AGN1}, 1/M_{\rm wdm} \}$ and $\textbf{X}$ are the 14 galaxy properties at $z=5$.

We find that the model has reasonably well approximated the true distribution of the data and show coverage tests of it in the Appendix \ref{ap:ct}. In Fig. \ref{fig:contour}, we show an example of the posterior distribution for one random galaxy of the test set. From this plot, we can see that the model can infer the value of $\Omega_{\rm m}$ and a bit of $A_{\rm SN1}$, but the other parameters are unconstrained. 

We test if this conclusion holds for other galaxies by taking random galaxies from the test set and drawing 10000 samples of their parameters using the normalizing flows.
We then compute the posterior mean and standard deviation from the parameters and show the results in Fig. \ref{fig:6_properties}. 

We find that our model is not able to infer the value of the parameters $\sigma_8$, $A_{\rm SN2}$, and $A_{\rm AGN1}$ (achieving $R^2<0.5$). On the other hand, the model can relatively accurately infer the value of $\Omega_{\rm m}$ ($R^2 = 0.9372$)  and, to a lesser extent, of $A_{\rm SN1}$ ($R^2 = 0.8037$). This is in agreement with the results of \cite{Paco2022APJ, Echeverri, Chaitanya}, even if here we use galaxies at $z=5$, the AGN parameter we vary is different to the one varied in those works, and we also vary the warm dark matter mass. We find that the model can also not infer the value of the WDM mass, and just return the input distribution. 

These results imply that the properties of individual galaxies at this redshift are not affected by the WDM mass or that its impact is smaller than that of other physical effects like cosmic variance. On the other hand, we find relatively tight constraints on $\Omega_\mathrm{m}$ even when varying the AGN parameters and dark matter physics: our model can infer the value of $\Omega_{\rm m}$ with an RMSE of $0.0410$, a mean relative error of $13.9\%$, and a coefficient of determination of $R^2=0.872$.

\begin{figure*}[]
\centering
    \includegraphics[width=1.0\columnwidth]{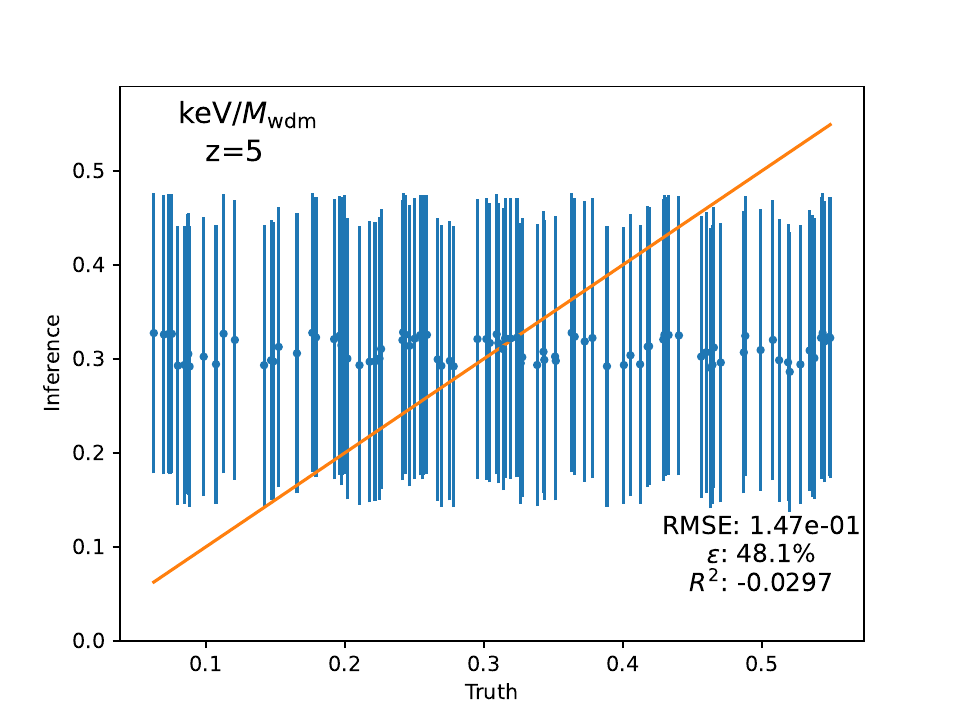}
    \includegraphics[width=1.0\columnwidth]{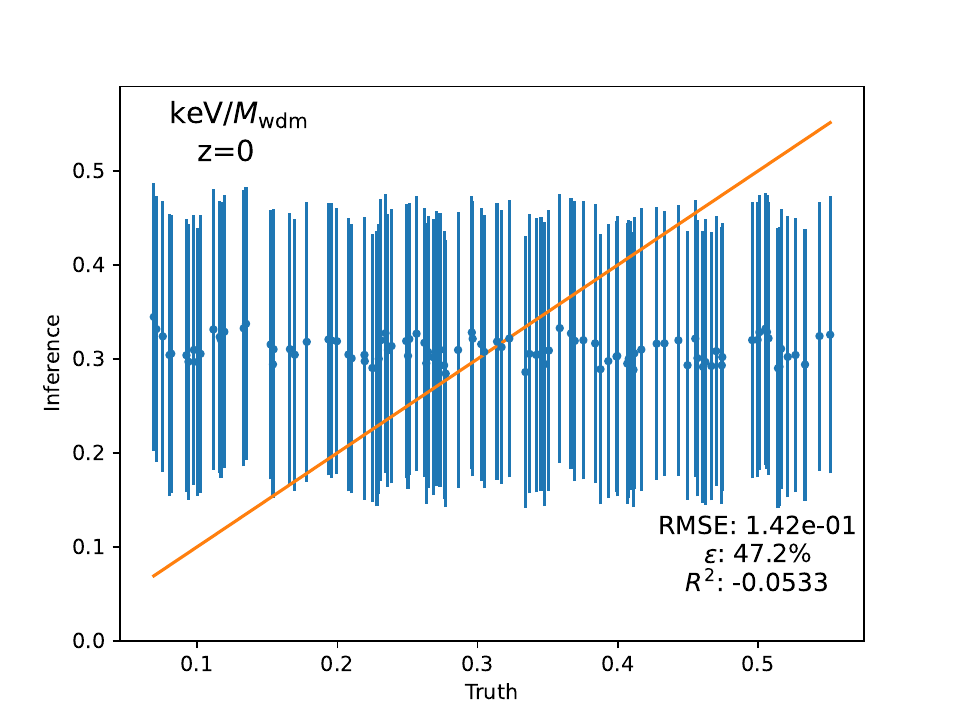}
     \caption{We have trained a model to learn the posterior distribution $p(\vec{\theta}|\mathbf{X'})$, where $\vec{\theta}$ is the WDM mass and $\mathbf{X'}$ is a 19-dimensional vector containing the 14 properties of individual galaxies at $z=5$ (left) or $z=0$ (right) plus the value of $\Omega_{\rm m}$, $\sigma_8$, $A_{\rm SN1}$, $A_{\rm SN2}$, $A_{\rm AGN1}$. Once the model is trained, we take one galaxy and draw samples from the posterior. We then calculate the posterior mean and its standard deviation from these samples. We have carried out this procedure for all \textit{small} galaxies (as defined in \ref{subsuc: gal_prop}) in a given simulation of the test set, and in the above plot, we show the mean of the posterior means and the mean of the posterior standard deviations. As can be seen, even in this scenario, the model cannot infer the WDM mass 
     (with mean relative error larger than $47.0\%$ in both cases), indicating that the lack of information is not due to degeneracies with other parameters.}
     \label{fig:WDM_mass}
\end{figure*}

\subsection{Degeneracies}
\label{subsec:small_gal}

One potential explanation of the above results is that the effects of WDM mass may be degenerate with those of cosmology and/or astrophysics. To investigate this, we have trained a new model to learn this posterior distribution $p(1/M_{\rm wdm}|\mathbf{X'})$, where $\mathbf{X'}$ is now a 19-dimensional vector containing the 14 galaxy properties plus the value of $\Omega_{\rm m}$, $\sigma_8$, $A_{\rm SN1}$, $A_{\rm SN2}$, $A_{\rm AGN1}$. Note that in this case, we are giving the model as input not only the galaxy properties but also the value of the cosmological and astrophysical parameters, so it can focus solely on determining the WDM mass.

The coverage test for this model shows that our model has learned a good approximation of the data distribution, as can be seen in Fig. \ref{fig:z=5_ct} of Appendix \ref{ap:ct}. Next, we take a random galaxy and draw 100 samples of the posterior. From them, we calculate the posterior mean and standard deviation. We then repeat this procedure for all galaxies in a simulation in the test set and compute
\begin{eqnarray}
\bar{\mu}=\frac{1}{N}\sum \mu_i\\
\bar{\sigma}=\frac{1}{N}\sum \sigma_i
\end{eqnarray}
which represents the mean prediction for the simulation. We show these results on the left panel of Fig. \ref{fig:WDM_mass}. 
As can be seen, even under this setup, the model is not able to infer the WDM mass from the properties of individual galaxies as the RMSE is larger than 0.1, the mean relative error is $48.1\%$ and the absolute value of $R^2$ drops below 0.1. We have repeated this test using the \textit{small galaxies} dataset as defined in \ref{subsuc: gal_prop}, reaching the same conclusions. This indicates that our results do not depend on galaxy selection, and it is likely a generic result.

\begin{figure*}
\centering
    \includegraphics[width=\columnwidth]{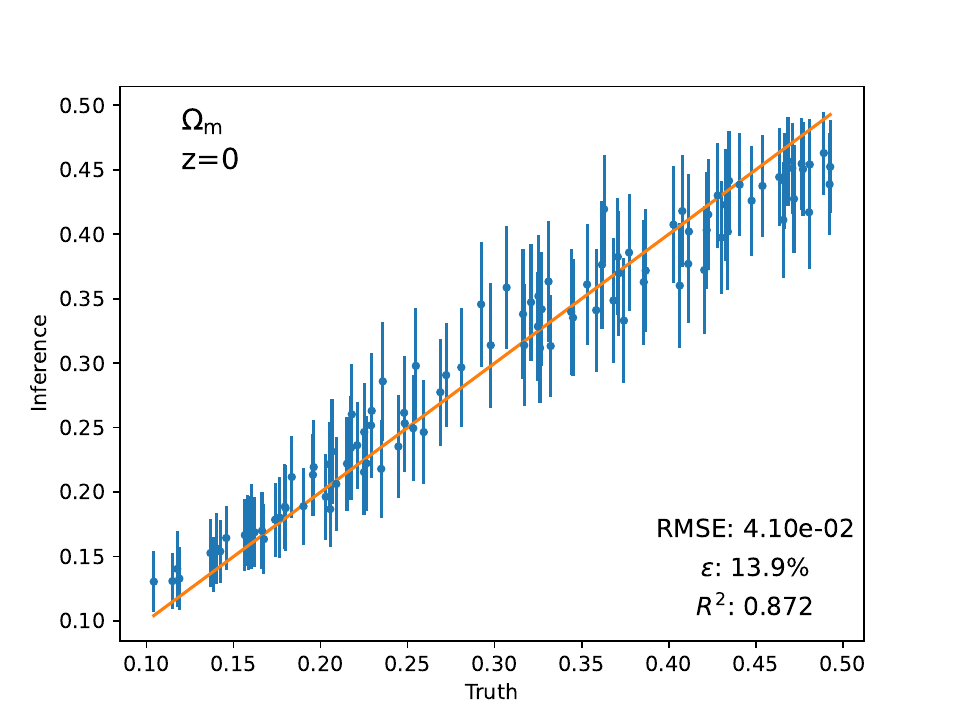}
    \includegraphics[width=\columnwidth]{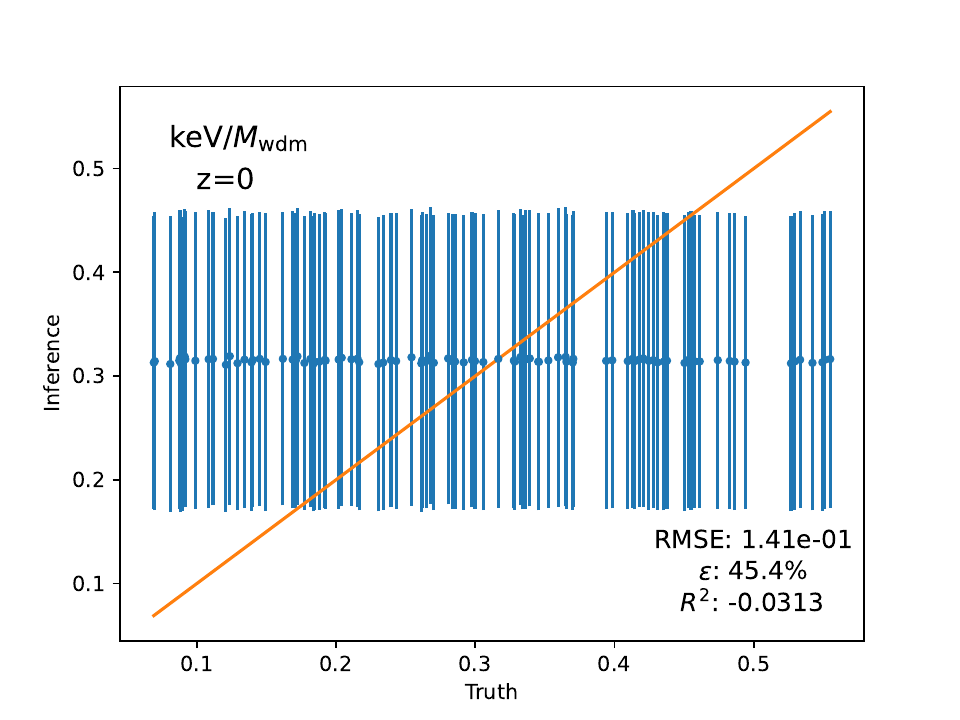}
     \caption{Same as Fig. \ref{fig:WDM_mass} but using all galaxies and learning the posterior of all parameters from the 14 galaxy properties at $z=0$. Our conclusions remain unchanged by using $z=0$ galaxies.}
     \label{fig:z=0}
\end{figure*}

\subsection{Redshift dependence}

We have trained two normalizing flow models using galaxies at $z=0$ to examine how redshift affects our results. The first model used as input the 14 galaxy properties of all the galaxies to infer the cosmological and astrophysical parameters and WDM mass, just like in Sec. \ref{subsec:all_gal}. The second model takes the 14 galaxy properties plus five cosmological and astrophysical parameters of small galaxies as input and learns the posterior distribution of the WDM mass, following \ref{subsec:small_gal}.

The coverage tests for these two models are shown in Fig. \ref{fig:z=0_ct} and \ref{fig:small_ct}, showing their reliability. As mentioned above, we generate samples of the posterior and calculate the means and standard deviations from a subset of random galaxies of the test set. In Fig. \ref{fig:z=0} we show the results for the model that only takes as input the galaxy properties. As can be seen, the model is not able to infer the WDM mass as the mean relative error comes to $45.4\%$, while it performs well when inferring $\Omega_{\rm m}$, achieving a mean relative error of $13.9\%$.

For the second model, we show the results on the right panel of Fig. \ref{fig:WDM_mass}. As can be seen, even in this case the model is not able to infer the warm dark matter mass. We note that naively, we would expect that the lower the redshift, the more difficult this task would be. This is because the impact of WDM is expected to be larger at higher redshift, where the suppression of power on small scales is expected to be cleaner since it has less time to be affected by non-linear gravitational evolution. 

\begin{figure}
    \includegraphics[width=\columnwidth]{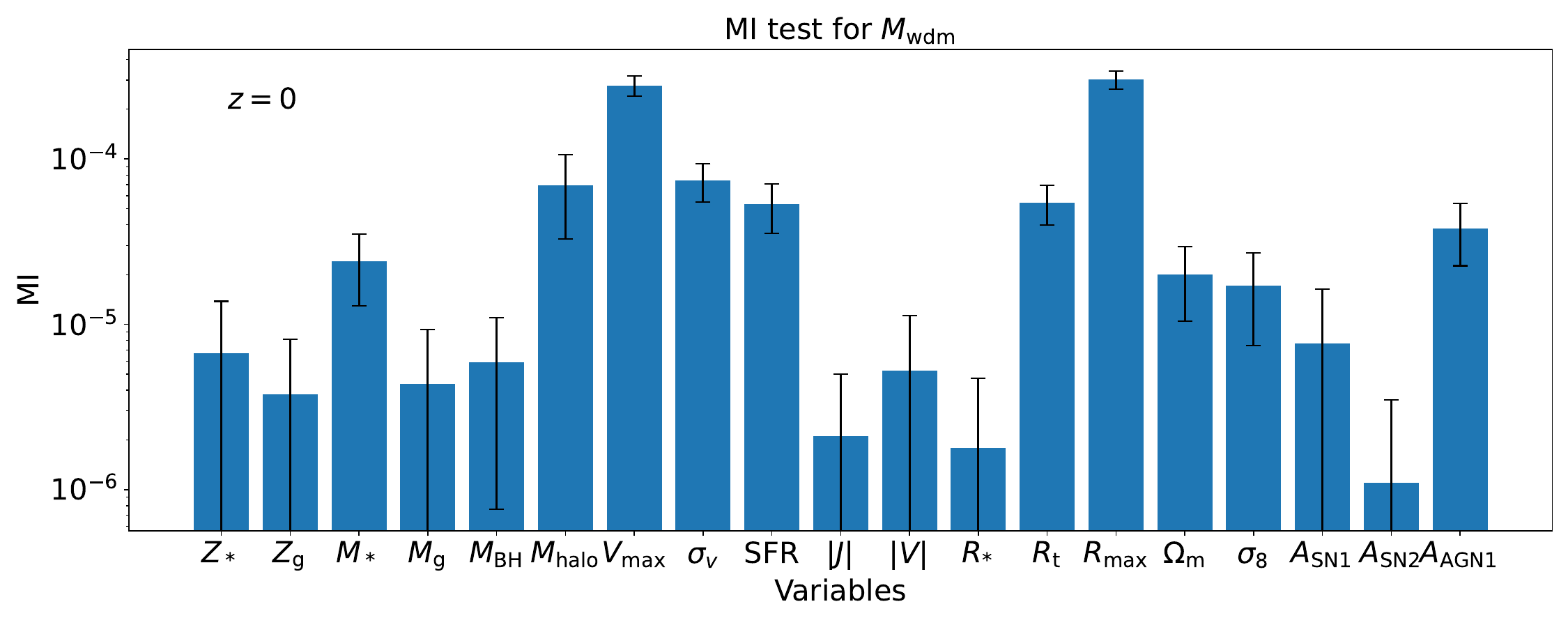}
    \includegraphics[width=\columnwidth]{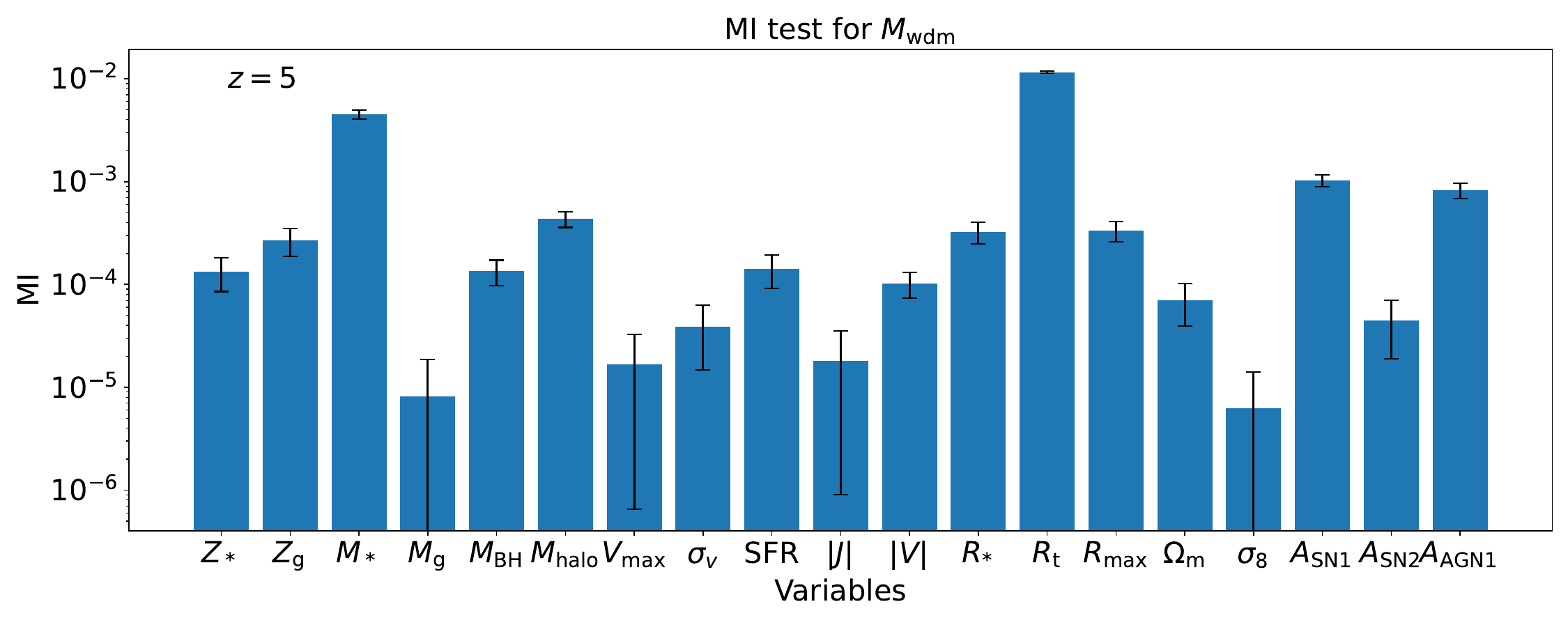}
     \caption{We have computed the mutual information between the WDM mass and the 14 galaxy properties at $z=0$ (top) and $z=5$ (bottom). As can be seen, we do not find any significant (non-linear) dependence between the WDM mass and any galaxy property. For reference, we have also computed the mutual information between $M_{\rm wdm}$ and the cosmological and astrophysical parameters that, by construction, are not correlated, finding their magnitude to be similar to the one of the galaxy properties.}
     \label{fig:MI}
\end{figure}

\subsection{Mutual Information}

Finally, we compute the mutual information between the different galaxy properties, cosmological parameters, astrophysical parameters, and $M_{\rm wdm}$. This allows us to determine the existence of a potential non-linear correlation between the galaxy properties and the WDM mass. We carry out this task using the \textit{GMM-MI} package \citep{Piras23GMM-MI}. This package first uses Gaussian mixture models to learn the joint distribution between two parameters and then uses that joint distribution to compute the mutual information.

We show the results of this calculation in Fig. \ref{fig:MI} at both $z=0$ (top) and $z=5$ (bottom). As can be seen, we find the values of the mutual information to be pretty small for all considered galaxy properties. This indicates the lack of a significant non-linear correlation between individual galaxy properties and the WDM mass. Note that we have also calculated the mutual information between $M_{\rm wdm}$ and the cosmological and astrophysical parameters, finding their values to have magnitudes similar to the ones of the galaxy properties. We emphasize that there should not be any correlation between $M_{\rm wdm}$ and the cosmological and astrophysical parameters, as they are all sampled simultaneously in the Sobol sequence.

We have also repeated this exercise but for $\Omega_{\rm m}$, finding much larger mutual information between it and some galaxy properties, like $V_{\rm max}$, that were identified as the most important in \cite{Paco2022APJ}. From this exercise, we can conclude that no individual galaxy property seems to exhibit a significant non-linear dependence on the WDM mass. This reinforces the conclusions we reached in the previous sections. 
%%%%%%%%%%%%%%%%%%%%%%%%%%%%%%%%%
%%%%%%% Conclusions %%%%%%%%%%%%%
%%%%%%%%%%%%%%%%%%%%%%%%%%%%%%%%%
\section{Conclusions} 
\label{sec:conclusions}
In light of recent work from \cite{Paco2022APJ}, who found a strong correlation between the value of $\Omega_{\rm m}$ and properties of individual galaxies, we investigate whether the changes induced by WDM (such as changes in the assembly history) leave signatures on the properties of individual galaxies.

Using a very large set of 1,024 state-of-the-art hydrodynamic simulations from the DREAMS project \citep{darkCAMELS}, we have trained normalizing flows to learn the posterior distribution $p(\vec{\theta}|\mathbf{X})$, where $\vec{\theta}$ is the value of cosmological, astrophysical, and the WDM mass parameters and $\mathbf{X}$ is the properties of individual galaxies. 

We find that our models cannot infer the value of individual galaxies' WDM mass, and the posterior distribution reflects the priors of the training set. This conclusion holds for galaxies at both $z=0$ and $z=5$. We find that providing the models with exact information about the cosmology and astrophysics of the considered galaxy does not help with the inference of the WDM mass. These results are found for all galaxies studied in this work, even small ones that perhaps are expected to be more affected by these effects. Finally, we have computed the mutual information between galaxy properties and the WDM mass, finding its amplitude very small in all cases.

These results indicate that any effect induced by WDM mass on simulated galaxies, if any, is much smaller than the intrinsic scatter from the population diversity itself. On the other hand, our models are able to accurately infer the value of $\Omega_{\rm m}$, in agreement with the results of \cite{Paco2022APJ,Echeverri, Chaitanya} even if the machine learning model is very different, the varied astrophysical parameters are different (the AGN parameter we vary in DREAMS is different to the one in CAMELS), and we have simulations with different WDM masses.

\begin{figure*}
\centering
\includegraphics[width=2\columnwidth]{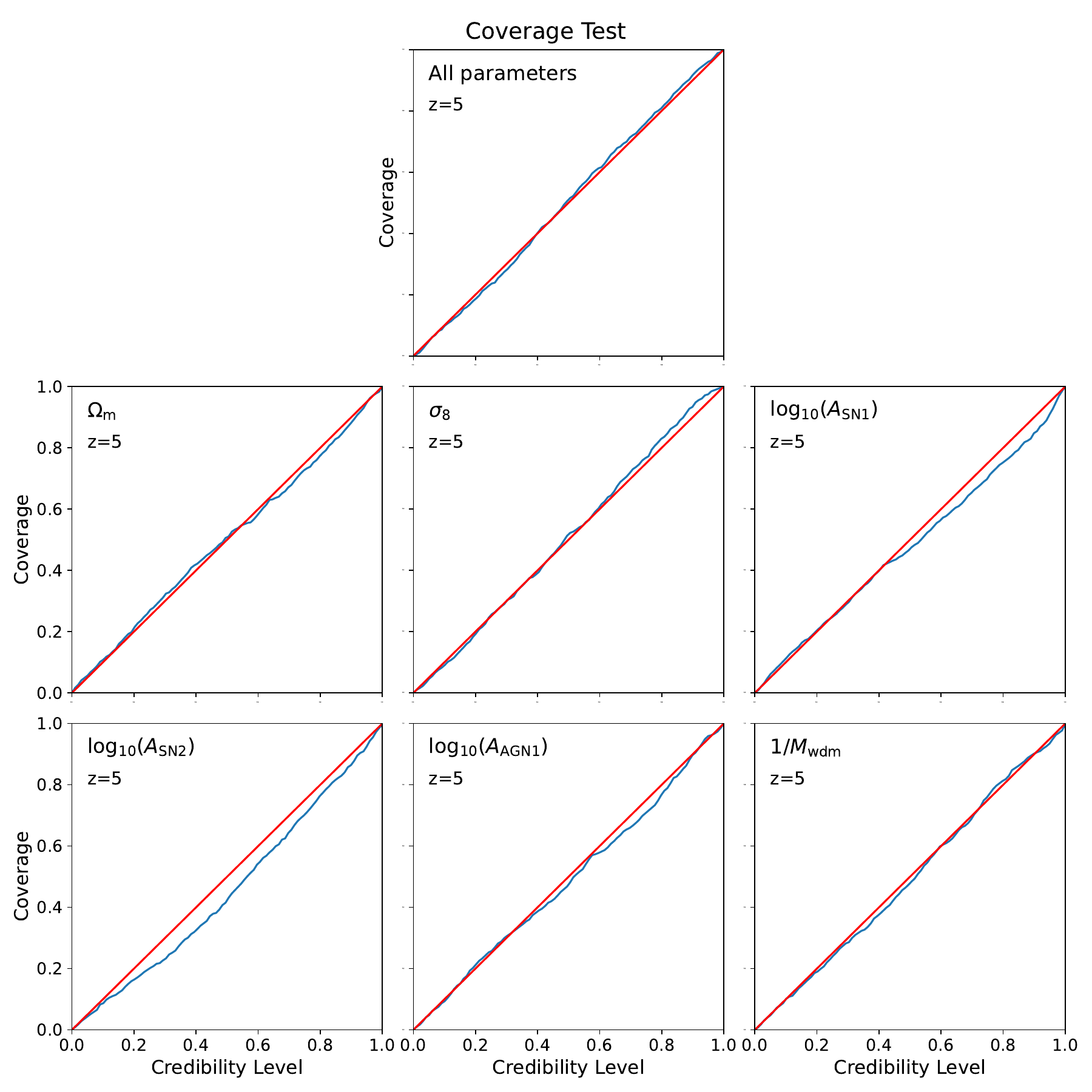}
\caption{This figure shows the coverage tests for the normalizing flow trained using as input the 14 galaxy properties at $z=5$ and as output the cosmological, astrophysical, and WDM mass parameters. As can be seen, our model has learned a good approximation of the underlying distribution.}
\label{fig:z=5_ct}
\end{figure*}

Finally, we would like to emphasize that there may be other ways to infer WDM masses for individual or small sets of galaxies. For instance, this analysis could be repeated with much smaller galaxies, where the impact of WDM mass may be more severe. In this case, higher-resolution simulations would be needed. One can also consider other properties, like stellar motions, not included in this analysis that may retain more information about WDM.

If individual galaxies cannot be used to infer WDM masses, it may still be possible, and expected, that sets of galaxies can. It would be interesting to quantify how much information can be extracted from an ensemble of them, e.g. using deep sets \citep{Bonny_2023} or GNNs \citep{PabloI, PabloII, PabloIII, Tri_2022, HelenI, NataliI}.

\section*{Acknowledgements}

The work of FVN is supported by the Simons Foundation. Yi-Fu Cai is supported in part by the National Key R\&D Program of China (2021YFC2203100), CAS Young Interdisciplinary Innovation Team (JCTD-2022-20), NSFC (12261131497), 111 Project (B23042), Fundamental Research Funds for Central Universities, CSC Innovation Talent Funds, USTC Fellowship for International Cooperation, USTC Research Funds of the Double First-Class Initiative. We acknowledge the use of computing clusters LINDA \& JUDY of the particle cosmology group at USTC. Arya Farahi was supported in part by the NSF AI Institute for Foundations of Machine Learning (IFML).

\appendix

\section{Coverage Test}
\label{ap:ct}

We performed a coverage test for every normalizing flow we have trained to assess its reliability. For this, we used the TARP method, presented in \citep{lemos2023arXiv}. If our model fits the data perfectly, the coverage should be equal to the credibility level. If the coverage deviates from the credibility level, our model may have a bias. If the curve is twisted into an ``S'' shape, it means an overestimation/underestimation.

\begin{figure*}
\centering
\includegraphics[width=\columnwidth]{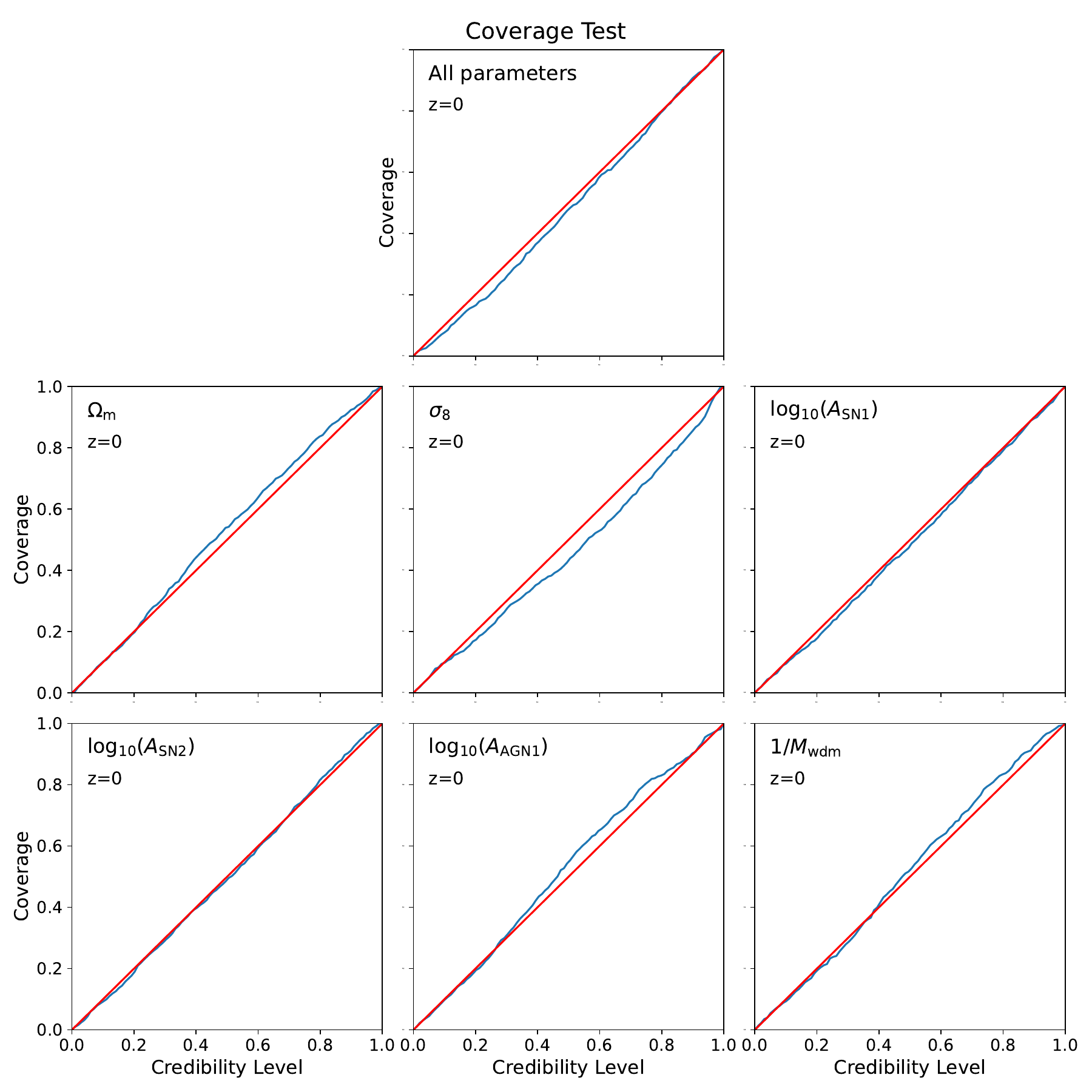}
\caption{Same as Fig. \ref{fig:z=5_ct} but for galaxy properties at $z=0$.}
\label{fig:z=0_ct}
\end{figure*}

\begin{figure*}
\centering
\includegraphics[width=0.45\columnwidth]{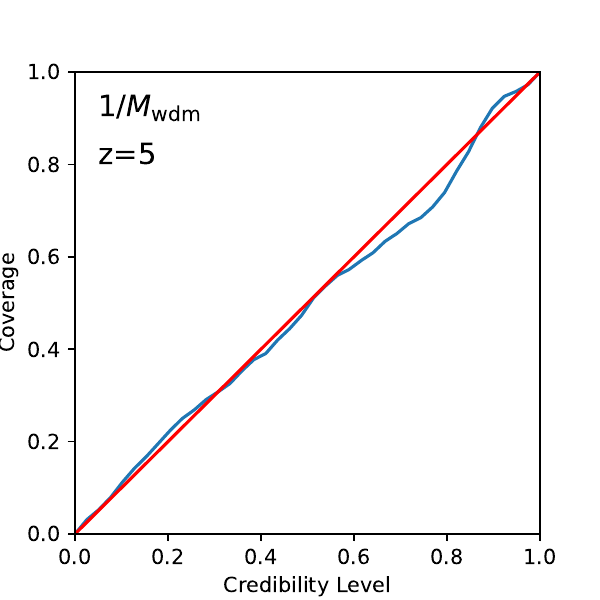}
\includegraphics[width=0.45\columnwidth]{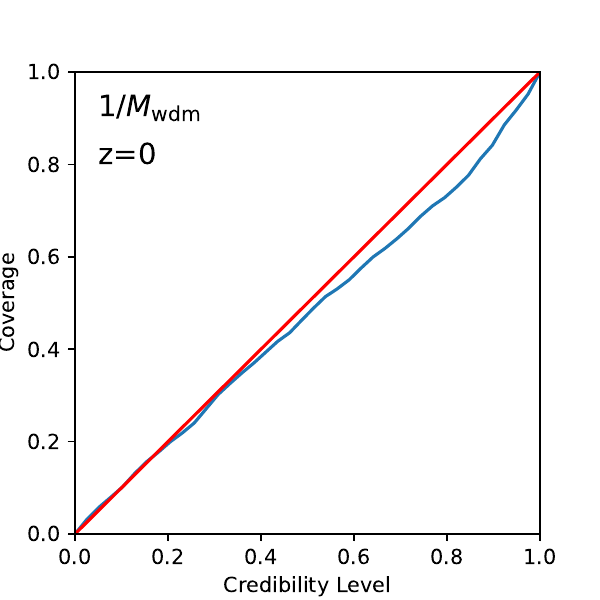}
\caption{Same as Figs. \ref{fig:z=5_ct} and \ref{fig:z=0_ct} but for small galaxies and using as input the 14 galaxy properties at $z=5$ (left) and $z=0$ (right) together with the value of $\Omega_{\rm m}$, $\sigma_8$, $A_{\rm SN1}$, $A_{\rm SN2}$, and $A_{\rm AGN1}$.}
\label{fig:small_ct}
\end{figure*}

For each test, we randomly select 2048 galaxies from the test set and generate 2048 samples from the posterior distribution. We also select 2048 random galaxies from the test set as true samples of the distribution. We note that in the cases where the posterior depends on several parameters, we have carried out the test for each individual parameter but also for all parameters at the same time. Fig. \ref{fig:z=5_ct},\ref{fig:z=0_ct},\ref{fig:small_ct} show the coverage tests for the different cases studied in this work.

\bibliography{reference}{}
\bibliographystyle{aasjournal}

\end{document}